# Avoiding quenching in superconducting rings

Autors: H. González-Jorge, D. González-Salgado, J. Peleteiro, E. Carballo, and G. Domarco

*Departamento de Física Aplicada, Universidad de Vigo, Facultad de Ciencias, Campus de Ourense, E-32004 Ourense, Spain*

**Abstract**

A procedure for avoiding the quenching associated to a previously reported method for inducing currents in superconducting rings [H. González-Jorge *et al.*, *Appl. Phys. Lett.* **81**, 4207 (2002)] is proposed. High persistent currents are induced in two textured YBCO rings (1536 and 2870 A) using a modified version of the original procedure that consist on a straightforward magnetic circuit made up of an iron core and a permanent magnet. The insertion of an iron core into the ring reduces the current to an extent sufficing to avoid quenching as the critical temperature is overcame and preserving thus the structures of the samples. Such a simple operation can be extended to other situations involving quenching (*e.g.* the use of superconducting coils).

**Keywords:** Quenching; superconducting rings; field cooling procedure

**PACS:** 74.25.Sv; 74.72.Bk; 74.81.Bd

---

Electronic mail: domarco@uvigo.es

In a previous work, an alternative method based on field cooling was developed with a view to inducting persistent currents in superconducting rings.[1] This method has the advantage that it allows a high current to be induced in a simple, convenient manner; however, it is subject to some shortcomings after the current is induced. Under these conditions, the most immediate choice for returning the sample to its normal state is by exceeding the critical temperature $T_C$, as the procedure by itself does not allow one to control the induced current. The use of the textures superconducting rings (*viz.* rings with a high critical current $I_C$) is subject to the risk of quenching, *i.e.* a large amount of heat is dissipated damaging –or even completely destroying– the structure of the material concerned.

In this work, we devised a straightforward procedure to avoid such quenching process. The procedure essentially relies on a magnetic interaction between a superconducting ring and an iron core. We also developed a variant of the inductive process that dispenses with the use of inductive coils.

The superconducting rings used in this work were two YBCO monocrystals supplied by the Institut für Material Physik (Göttingen, Germany) that were obtained from bulk samples of textured YBCO,[2] both samples were identical in composition and dimensions (36×36×15 mm$^3$). Two concentric holes of 10 and 20 mm in diameter, respectively, were made in each to obtain respective rings. The initial critical current in the superconducting rings, $I_C^i$ was measured by using a previously reported inductive method.[3,4] As can be seen from Table I, there was a substantial difference in critical current between the two samples; this was a result of defective crystal growth in the sample with the lower critical current (Y1).

Table I. Critical current for samples Y1 and Y2 before $I_C^i$ and after $I_C^f$, the experiment; maximum current $I_M$ that can be induced in a ring with the dimensions of Y1 and Y2; magnetic field $B_H$ and current $I_H$ induced in the samples; and reduced current $I_{red}$ circulating through the samples upon insertion of the iron core.

| Sample | $I_C^i$ (A) | $I_M$ (A) | $B_H$ (T) | $I_H$ (A) | $I_{red}$ (A) | $I_C^f$ (A) |
|---|---|---|---|---|---|---|
| Y1 | 1542 | 16037 | 0.091 | 1536 | 26 | 1539 |
| Y2 | 2877 | 16037 | 0.170 | 2870 | 30 | 2873 |

In addition to the typical current induction procedures (*viz.* field cooling and zero field cooling),[5-7] one can use the modified version of the field cooling[1] method proposed by us as it considerably simplifies the induction of critical currents. Essentially, magnetizing a ferromagnetic core by means of a small inductive coil allows one to induce a high critical current in a superconducting ring. In the modified procedure, the coil is replaced with a permanent magnet (see Fig. 1.), which further simplifies the process as it allows the current to be induced without the need for a coil or power supply. To our knowledge, this is the first method where current is induced simply by using energy from magnetic material, which makes it especially elegant. The specific steps of this new version of our method are similar to those of the original one and are described in detail in the Annex.

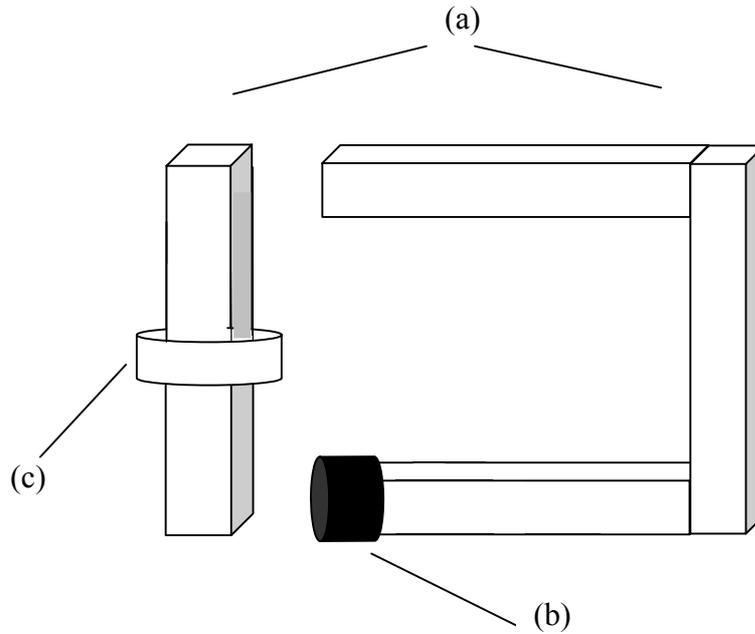

Figure 1. Experimental set-up for the induction of persistent current in superconducting rings. *(a)* Ferromagnetic core. *(b)* Nd-Fe-B magnet. *(c)* Superconducting rings.

Based on the data in Table I, one can induce maximum persistent currents [$I_M$, see section (2) in the Annex] roughly one order of magnitude higher than the critical currents used in this work $I_C^i$, ensuring saturation as a result. The current induced in the superconductor, $I_H$, was calculated from the magnetic field $B_H$ measured with a cryogenic Hall probe; the results are also shown in Table I. Once the current is induced, quenching is avoided by manipulating the rings before they are removed from the liquid nitrogen. The steps involved and their physical foundation are as follows:

(1) A new iron core is inserted into the superconductor as depicted in Fig. 2. On passage through the superconducting sample, the core is magnetized and the total flux across it is increased as a result. This decreases the current in the superconductor so that the magnetic flux $\phi$ remains constant. The current that circulates along the superconducting ring under these new conditions can be readily calculated from $\oint \vec{H} \, \vec{dl} = I$, where $H$ is the magnetic field produced by the superconductor, $I$ the current circulating through it, and $l$ the length of the ferromagnetic core. Under these conditions, the magnetic field in the circuit can be expressed as:

$$B = \frac{\mu_0 (1+\chi) I}{L} \qquad [1]$$

where $\mu_0$ is the magnetic permeability, $\chi$ the susceptibility of the ferromagnetic core, and $L$ (41.3 cm) the length of the magnetic circuit. By making $B_H = B$ in this equation, and taking into account the magnetic properties of the core (see Fig. 3a), one can obtain an estimate of the reduced current $I_{red}$ that will circulate through the ring upon insertion of the iron core. The results obtained for both rings are also shown in Table I; as can be seen, the current intensity circulating through each ring was substantially decreased.

At this point, the samples can be removed from the liquid nitrogen without the risk of damaging their structure with the increase in temperature. The efficiency of the method can be checked by performing a new measurement of the critical current $I_C^f$ of the samples. The fact that $I_C^i$ was virtually identical with $I_C^f$ testifies to the effectiveness of the proposed method.

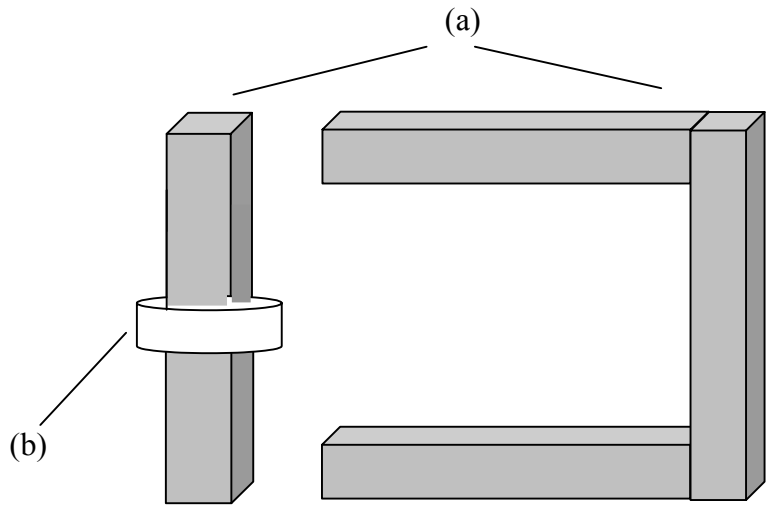

Figure 2. Experimental set-up for reducing persistent current in superconducting rings. *(a)* Ferromagnetic core. *(b)* Superconducting ring.

One could think about the possibility of using the magnetic induction circuit described in the Annex to avoid quenching. This is misleading, however, as, based on the magnitude of the inducting field, which exceeds one magnitude order the field generated by the currents, the critical current would be induced in the opposite direction.

Although the proposed procedure complements the inductive method previously reported by us, it could also be applicable to some quenching situations. The earliest procedures developed to avoid quenching were reported in the 1960s.[8] One of the most common situations in this context is that where a superconducting coil is disconnected form the power supply and short-circuited, so the energy dissipated when $T_C$ is exceeded must be internally absorbed. In these conditions, the coil is usually protected with a heat cooler (usually cooper made).[9] Under these conditions, the above-described procedure could be used to replace the cooper coolers as the transition to the normal state takes place at low current. Additional uses of interest remain to be developed.

**Acknowledgements**

The authors wish to thank Professors Claudio A. Cerdeiriña and Juan Faílde for their kind cooperation.

**Annex: Inducting current in superconducting rings by using a magnetic circuit consisting of a ferromagnetic core and a permanent magnet**

The current induction procedure introduced in this work involves the following steps:

(1) The magnetic circuit of Fig. 1 is assembled at room temperature. The circuit consists of an iron core and Nd-Fe-B magnet. The magnetic characteristics of the two elements are illustrated in Fig. 3. The iron core is passed through the superconducting ring.

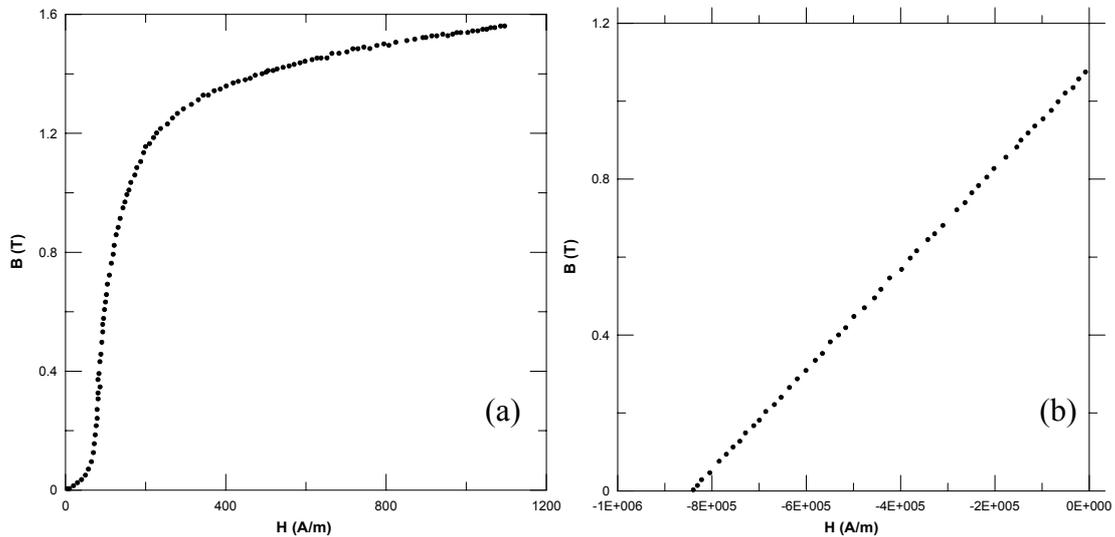

Figure 3. *(a)* Hysteresis curve for the first quadrant of the iron core. *(b)* Demagnetization curve for the second quadrant of an Nd-Fe-B permanent magnet.

(2) The fundamental laws of electromagnetism provide an estimate of the maximum current $I_M$ that can be induced with the magnetic circuit described in (1). Based on the experimental set-up used, $\oint \vec{H} \cdot \vec{dl} = NI$, where $H$ is the total magnetic field and $l$ the length of the magnetic circuit. In our case, $NI = 0$ as there is no coil around the core, so $\oint \vec{H} \cdot \vec{dl} = \left[ \int \vec{H} \cdot \vec{dl} \right]_{MAGNET} + \left[ \int \vec{H} \cdot \vec{dl} \right]_{CORE}$. In calculating the magnetic field in the circuit, one must consider the cross-sectional area and length of the core (1.12 cm² and 36 cm) and the magnet (1.77 cm² and 1.9 cm), respectively. Based on their magnetic characteristics, the resulting magnetic field in the circuit will be 0.95 T. As can be seen in Fig. 3a, the result was the magnetic field of the magnet when the demagnetizing field approached zero.

A persistent current circulating through a superconducting ring produced a magnetic field $B_H$ at the mid-point of its central axis that is given by:

$$B_H = \frac{\mu_0 I}{2\sqrt{(d/2)^2 + r^2}} \qquad [2]$$

where $d$ and $r$ are the length and mean radius, respectively, of the superconducting. In the superconducting state, the magnetic flux ϕ across the hole in the superconducting ring will be constant. Accordingly, $B_H$ will equal the magnetic field of the core (0.95 T) and the maximum current $I_M$ that can be induced for the ring dimensions ($d$ = 15 mm and $r$ = 7.5 mm in our case) will be obtained.

(3) The sample is cooled below $T_C$. Then, the magnetic circuit is opened and the superconducting ring, with its induced current, is removed from it.

**Bibliografía**

[1] H. González-Jorge, J. Peleteiro, E. Carballo, L. Romaní, and G. Domarco, Appl. Phys. Lett. **81**, 4207 (2002).

[2] Ch. Jooss, B. Bringmann, M. P. Delamare, H. Walter, A. Leenders, and H. C. Freyhardt, Supercond. Sci. Technol. **14**, 260 (2001).

[3] A. Díaz, A. Pomar, G. Domarco, C. Torrón, J. Maza, and Felix Vidal, Physica C **215**, 105 (1993).

[4] G. Domarco, A. Díaz, A: Pomar, O. Cabeza, C. Torrón, J. A. Veira, J. Maza, and F. Vidal, in: Superconductivity, ICMAS'92, Eds. C. W. Chu and J. Fink (IITT-Internacional Gournay sur Marne, 1992) p. 105.

[5] J. Jung, I. Isaac, and M. A.-K- Mohamed, Phys. Rev. B **48**, 7526 (1993).

[6] I. Isaac, J. Jung, M. Murakami, S. Tanaka, M. A.-K. Mohamed, and L. Friedrich, Phys. Rev. B **51**, 11806 (1995).

[7] I. Isaac and J. Jung, Phys Rev. B **55**, 8564 (1997).

[8] P. F. Smith, Rev. Sci. Instrum. **34**, 368 (1963).

[9] F. Rodríguez-Mateos, B. Szeless, and F. Calvone, LHC Project Report **48**,23 September 1996.